\documentstyle[12pt]{article}\textheight 230mm\textwidth 150mm
\newcommand{\be}{\begin{eqnarray}}
\newcommand{\ee}{\end{eqnarray}}
\newcommand{\eel}[1]{\label{#1}\end{eqnarray}}
\newcommand{\r}[1]{(\ref{e:#1})}
\newcommand{\ben}{\begin{enumerate}}
\newcommand{\een}{\end{enumerate}}
\newcommand{\beq}{\begin{quote}}
\newcommand{\eq}{\end{quote}}
\newcommand{\bp}{{\bf p}}
\newcommand{\vb}{{\cal h}}
\newcommand{\hb}{{\cal i}}

\newcommand{\ra}{{\rightarrow}}
\newcommand{\nn}{\nonumber}
\newcommand{\eg}{{\em e.g.\ }}
\newcommand{\ie}{{\em i.e.\ }}
\newcommand{\al}{\alpha}
\newcommand{\ga}{{\gamma}}
\newcommand{\la}{{\lambda}}

\newcommand{\del}{{\delta}}
\newcommand{\Om}{\Omega}
\newcommand{\pet}{{\cal P}}
\newcommand{\bata}{\bar{\eta}}
\newcommand{\bapet}{\bar{\pet}}
\newcommand{\bac}{\bar{c}}
\newcommand{\bak}{\bar{k}}
\newcommand{\dagg}{^{\dag}}

\newcommand{\halv}{\frac{1}{2}}

\begin{document}
\begin{titlepage}
\noindent
G\"{o}teborg ITP 93-20\\
August 1993\\
\vspace*{30 mm}
\begin{center}{\LARGE\bf BRST quantization of relativistic particles on inner
product
spaces\footnote{Talk presented at  the second workshop on "Constraint Theory
and Quantization
Methods", Montepulciano, Italy, June 28 -	July 1,
1993.}}\end{center}\begin{center}
\vspace*{12 mm}

\begin{center}Robert Marnelius \\
\vspace*{7 mm}
{\sl Institute of Theoretical Physics\\
Chalmers University of Technology\\
S-412 96  G\"{o}teborg, Sweden}\end{center}
\vspace*{25 mm}
\begin{abstract}
Recent results of BRST quantization on inner product
spaces are reviewed. It is shown how relativistic particle models may be
quantized with finite
norms and that the  relation between the operator method and the
conventional path integral treatments is nontrivial.
\end{abstract}\end{center}\end{titlepage}

\setcounter{page}{1}

BRST quantization within the operator formulation has the following
ingredients: One starts with a
nondegenerate state space, $\Om$, and projects out a physical subspace by the
condition $Q\Om_{ph}=0$
where $Q$ is the BRST charge operator which must be nilpotent $Q^2=0$. Then one
makes $\Om_{ph}$
nondegenerate by dividing out $Q\Om$ from $\Om_{ph}$. When this procedure is
applied to relativistic
particle models one usually derives covariant field equations (see \eg
\cite{MU}). However, when one
applies the corresponding path integral formalism then one usually derives
propagators
\cite{Teit,PRi,Gov}. It is therefore pertinent to ask how the two formalisms
are related.
Another question one may ask is whether or not it is possible to check
unitarity of
relativistic particle models, \ie is it possible to calculate norms of physical
states and check
whether or not  they are positive?  In this talk I shall show that the answer
to these questions is
satisfactory provided  the original state space $\Om$ is  chosen to be an inner
product space.

In \cite{HM} a general framework for BRST quantization was proposed and
analysed. In this framework
the original state space $\Om$ needs not be an inner product space. However,
when it is not, one
is forced to consider its dual state space $\Om'$. (One needs the finite
bilinear forms, $|'\vb
u'|u\hb|<\infty$ \cite{HM}.) This framework was inspired by conformal field
theory methods in string
theory where the zero modes are treated in this asymmetric fashion. The BRST
condition leads here to
two different physical state spaces: $Q|ph\hb=0$ and $'\vb ph|Q=0$ where
$|ph\hb\in\Om_{ph}\subset\Om$
and $|ph\hb'\in\Om'_{ph}\subset\Om'$. These conditions may be solved by a
bigrading \cite{HM,Bi}
which means that $Q|ph\hb=0$ may be replaced by \be &&\del|ph\hb=d|ph\hb=0
\eel{e:1}
where
\be
&&Q=\del+ d,\;\;\;\del^2=d^2=[\del, d]_+=0
\eel{e:2}
However, $'\vb ph|Q=0$ implies
\be
&&\del\dagg|ph\hb'=d\dagg|ph\hb'=0
\eel{e:3}
which means that
\be
&&\Om'_{ph}\neq\Om_{ph}
\eel{e:4}
when $d\dagg\neq\del,d$. The problem is therefore how to make sense of
$\Om'_{ph}$ when \r{4} is
valid. In several models it was possible to interpret $'\vb ph|ph\hb$ as a
finite wave function
\cite{MU}. However, no norms are derivable although an effort was made in
\cite{MO}.

The above framework is considerably simplified when $\Om$ is required to be an
inner product space.
In this case $\Om'=\Om$ which implies $\Om'_{ph}=\Om_{ph}$.  This
simplification is, however, only
possible under certain conditions. They are (see \eg \cite{Bi})
\ben
\item Nontrivial states in $\Om_{ph}$ has ghost number zero.
\item The number of (hermitian) gauge generators (constraints) must be even.
\item $Q$ must be possible to decompose as
\be
&&Q=\del+\del\dagg,\;\;\;\del^2=[\del, \del\dagg]_+=0
\eel{e:5}
\een
In order to see what the last two conditions require I consider a general
bosonic gauge theory. The
BRST charge may then be written in the BFV form \cite{BV}
\be
&&Q=\psi_a\eta^a-\frac{1}{2}iU_{bc}^{\;\;a}\pet_a
\eta^b\eta^c-\frac{1}{2}iU_{ab}^{\;\;b}\eta^a + \bapet_a\pi^a
\eel{e:6}
where $\psi_a,\;a=1,\ldots,m$ are hermitian bosonic gauge generators
(constraints) satisfying the Lie algebra
\be
&[\psi_a, \psi_b]_{-}=iU_{ab}^{\;\;c}\psi_c
\eel{e:7}
where $U_{ab}^{\;\;c}$ are the structure constants.  $\eta^a$ and $\bata^a$ are
ghost and antighost
and $\pi_a$ is the conjugate momentum to the Lagrange multiplier $v^a$. They
are hermitian and
satisfy the  algebra (nontrivial part)
\be
&&[\eta^a, \pet_b]_+=[\bata^a,\bapet_b]_+=\del^a_b,\;\;\;[\pi_a,
v^b]_-=-i\del^b_a
\eel{e:8}
We notice that the inclusion of dynamical Lagrange multipliers will always
ensure  that the second
condition above is satisfied, since we have an even number of constraints:
$\psi_a=0,$ $\pi_a=0$.
Recently \cite{Simple,Gauge}, I have  shown that it is always possible to
decompose the BRST
charge \r{6} according to \r{5} which means that also condition three above is
satisfied  for
the BRST charge \r{6}. The general procedure to do this is to look for a
unitary transformation which
preserves the ghost number:
\be
&&\eta, \pet, \bata, \bapet, \ldots\longrightarrow \eta', \pet', \bata',
\bapet', \ldots
\eel{e:9}
Typically it involves the Lagrange multipliers \cite{Simple} or a gauge fixing
operator $\chi^a$
\cite{Gauge}. Define the complex ghosts $c^a$ and $k_a$ by
\be
&&c^a\equiv\halv(\eta'^ a-i\bapet'^a), \; \;\;k_a\equiv\pet'_a-i\bata'_a
\eel{e:10}
They satisfy
\be
&&[k_a, c^{\dag b}]_+=\del_a^b
\eel{e:11}
Now  if
we define $\del$ by
\be
&&\del\equiv[c^{\dag a}k_a, Q]
\eel{e:12}
then $Q=\del+\del\dagg$. However, this expression is only nilpotent if also
\be
&&[k_a\dagg c^a, \del]=0
\eel{e:13}
This condition selects possible unitary transformations \r{9}. The solutions
found in
\cite{Simple,Gauge} all had the form
\be
&&\del=c^{\dag a}\phi_a=\phi'_ac^{\dag a}
\eel{e:14}
where $\phi_a^{(')}$ are nonhermitian operators which satisfy the same algebra
as $\psi_a$ \ie
\be
&[\phi_a, \phi_b]_{-}=iU_{ab}^{\;\;c}\phi_c
\eel{e:15}
(In some cases $\phi_a$ may be chosen  to be abelian \cite{Gauge}.) The BRST
condition $Q|ph\hb=0$ may
now be solved by a bigrading which yields
\be
&&\del|ph\hb=0,\;\;\;\del\dagg|ph\hb=0
\eel{e:16}
One may show that all solutions except possibly some zero norm states also are
solutions of
\be
&&c^a|ph\hb=\phi_a|ph\hb=0
\eel{e:17}
or
\be
&&c^{\dag a}|ph\hb={\phi'}^{\dag}_a|ph\hb=0
\eel{e:18}
Both these sets of equations have nontrivial solutions only for ghost number
zero according to the
analysis in \cite{Aux}: The nontrivial solutions of \r{17} satisfy
\be
&&c^a|ph\hb=k_a|ph\hb=0
\eel{e:19}
while \r{18} implies
\be
&&c^{\dag a}|ph\hb=k\dagg_a|ph\hb=0
\eel{e:20}
both of which requires $|ph\hb$ to have ghost number zero. Notice that \r{17}
and \r{18} are
consistency conditions to \r{19} and \r{20} since $\phi_a=[Q, k_a]_+$ and
${\phi'}^{\dag}_a=[Q,
k\dagg_a]_+$. In \cite{Simple,Gauge} general solutions of \r{17} and \r{18}
were derived all of the
form \be
&&e^{\al[\rho, Q]_+}|\Phi\hb
\eel{e:21}
where $\al$ is a real constant different from zero and where $|\Phi\hb$ is a
simple BRST invariant
state with ghost number zero. Two cases have been found:
\ben
\item)\hspace{1cm} $\rho=\pet_av^a$ with
$\eta^a|\Phi\hb=\bata^a|\Phi\hb=\pi_a|\Phi\hb=0$
\item)\hspace{1cm} $\rho=\bata_a\chi^a$ with
$\pet^a|\Phi\hb=\bapet^a|\Phi\hb=(\psi_a+\halv i
U_{ab}^{\;\;b})|\Phi\hb=0$
\een
where $\chi^a$ is a hermitian gauge fixing operator which must be such that
$[\chi^a, \psi_b]$ has
an inverse. These general solutions are obtained in a purely algebraic way.
They are therefore
formal. However, if the quantization is such that $\Om$ is an inner product
space then they must
also belong to an inner product space.  Thus, although \r{21} implies
\be
&&|ph\hb=|\Phi\hb+Q|\cdot\hb
\eel{e:22}
the states $Q|\cdot\hb$ cannot be divided out since $|\Phi\hb$ is not an inner
product state while
$|ph\hb$ must be. The second case above provides for a way to make the
solutions of a
Dirac quantization consistent with an inner product space since
 $|\Phi\hb$  is a solution to a Dirac
quantization there and
$\vb\Phi| e^{\al[\rho, Q]_+}|\Phi\hb$
is finite.

In order to find out what quantization rules have to be used in order for
\r{21} to belong to an
inner product space I consider first the trivial  case when the gauge symmetry
is abelian and
$\psi_a=p_a$ where $p_a$ is a conjugate momentum operator to some coordinate
operator $x^a$
\cite{Propa}. The BRST charge is here
\be &&Q=p_a\eta^a+\bapet^a\pi_a=\del+\del\dagg
\eel{e:23}
where $\del=c^{\dag a}\phi_a$ where in turn
\be
&&c^a=\halv(\eta^a-i\bapet^a),\;\;\;\phi_a=p_a-i\pi_a
\eel{e:24}
(Actually this case has been solved in its full generality in \cite{Slav}.) If
\eg the conditions
\r{16} allow for \r{17} then the solutions are
\be
&&|ph\hb=c^{\dag a}|\cdot\hb,\;\phi\dagg_a|\cdot\hb
\eel{e:25}
which all are zero norm states except for the vacuum state $|0\hb$ which
satisfies
\be
&&k_a|0\hb=\xi^a|0\hb=0
\eel{e:26}
where
\be
&&k_a\equiv\pet_a-i\bata_a,\;\;\;\xi^a\equiv\halv(ix^a-v^a),\;\;\;[\xi^a,
\phi\dagg_b]_-=\del^a_b
\eel{e:27}
The diagonal basis of this Fock space is spanned by
\be
&&a_a=\halv(\xi_a+\phi_a),\;\;\;b_a=\halv(\xi_a-\phi_a)
\eel{e:28}
where $a\dagg_a$ spans positive metric states and $b\dagg_a$ indefinite ones.
Obviously the
quantization should be such that  the unphysical states have a basis consisting
of half of positive
metric states and half of indefinite ones.

In order to understand what this result means for spectral bases let me
consider the hermitian
coordinate and momentum operators $Q$ and $P$ satisfying
\be
&&[Q, P]_-=i
\eel{e:29}
Real spectral decompositions requires positive definite states since
\be
&&Q|q\hb=q|q\hb,\;\;\;\vb q|q'\hb=\del(q-q')
\eel{e:30}
implies
\be
&&\vb\phi|\phi\hb=\int dq|\phi(q)|^2>0
\eel{e:31}
There is also the possibility to use imaginary spectral
decompositions as  Pauli showed in \cite{Pauli}:
\be
&&Q|iq\hb=iq|q\hb,\;\;\;\vb iq|iq'\hb=\del(q-q'),\;\;\;\vb -iq|=(|iq\hb)\dagg
\eel{e:32}
which requires indefinite metric states
\be
&&\vb\phi|\phi\hb=\int dq\phi^*(-q)\phi(q)=\pm C,\;\;\;C>0
\eel{e:33}
where the sign depends on the parity $\phi(-q)=\pm\phi(q)$.

Consider now the general solution \r{21} with $\rho=\pet_av^a$ for the trivial
case \r{23} \ie
\be
&&|ph\hb=e^{\al[\rho, Q]_+}|\Phi\hb=e^{\al(p_av^a+i\pet_a\bapet^a)}|\Phi\hb
\eel{e:34}
It has the norm
\be
&&\vb ph|ph\hb=\vb\Phi|e^{2\al[\rho, Q]_+}|\Phi\hb=\vb\phi|\,_{\pi}\vb0|e^{2\al
p_av^a}|0\hb_{\pi}|\phi\hb \eel{e:35}
The condition that this is a finite expression leads to the quantization rule:
\beq
{\sl Rule 1}: Lagrange multipliers must be quantized with opposite metric
states to the unphysical
variables which the gauge generator $\psi_a$ eliminates.
\eq
 For \r{35} it implies
\be
&&\vb ph|ph\hb=\int d^mp d^mv e^{i\al
p_av^a}|\phi(p)|^2=\frac{1}{\al^m}|\phi(0)|^2
\eel{e:36}
It also implies that there must be equally many positive as indefinite metric
oscillators in
accordance with the result above. Thus, this rule is at least true for the
trivial case.

Reconsider  the trivial case \r{23}
\be
&&Q=p_a\eta^a+\bapet^a\pi_a
\eel{e:37}
Obviously one may also make the choice to interpret $\eta^a$ as gauge generator
and $\bata_a$ as
Lagrange multiplier, and $p_a$ and $\pi_a$ as bosonic ghost and antighost
respectively. Finiteness of
the norms imply then the quantization rule:
\beq
{\sl Rule 2:} Bosonic ghosts and antighosts must be quantized with opposite
metric states.
\eq
In addition finiteness requires sometimes that the range of the Lagrange
multipliers is restricted
to the group manifold.
(Notice that $\vb ph|ph\hb=\vb\Phi|e^{[\rho,
Q]}|\Phi\hb=\vb\Phi|e^{\psi_av^a\cdots}|\Phi\hb$ where
$e^{\psi_av^a}$ is a finite group transformation.)

The restriction to inner product spaces leads obviously to severe restrictions
of the
allowed quantization. One may therefore question the possibility to quantize
relativistic particles
and strings in a manifestly Lorentz covariant way on inner product spaces. The
issue here is whether
or not one may represent hermitian manifestly Lorentz covariant coordinate and
momentum operators
$X^{\mu}$ and $P^{\mu}$ satisfying $[X^{\mu},P^{\mu}]_-=i\eta^{\mu\nu}$ in a
satisfactory way.
The possible representations are given in the table below \cite{Gen}

\begin{tabular}{|l|l|lr|}\hline
Spectra&Lorentz covariant basis&Hermitian inner products&\\ \hline \hline
real&no, positive&yes&\\ \hline
imaginary $x^0,\;p^0$&yes, indefinite&yes&\\ \hline
real&yes, imaginary&no&\\ \hline
\end{tabular}\\ \\
The last possibility is unsatisfactory and will not be considered. Thus, we
have a choice between
a manifestly covariant spectrum or basis. (Off-shell states are described by
Euclidean spectra and a
Lorentz covariant basis as we shall see.)

Consider first the free spinless particle \cite{Propa}. It is described by the
mass shell condition.
In the corresponding BRST formulation it is described by the BRST charge
operator
\be
&&Q=\halv(P^2+m^2)\eta+\pi\bapet=c\dagg\phi+\phi\dagg c
\eel{e:38}
where
\be
&&\phi=\halv(P^2+m^2)-i\pi,\;\;\;c=\halv(\eta-i\bapet)
\eel{e:39}
Thus, the BRST condition $Q|ph\hb=0$ leads to the possible solutions
\be
&&|ph\hb_{\pm}=e^{\pm\halv(P^2+m^2)v\pm i\pet\bapet}|\Phi\hb
\eel{e:40}
where $|\Phi\hb$ satisfies
\be
&&\pi|\Phi\hb=\eta|\Phi\hb=\bata|\Phi\hb=0
\eel{e:41}
Its formal norm is
\be
&&_{\pm}\vb ph|ph\hb_{\pm}\propto\vb\phi|e^{\pm(P^2+m^2)v}|\phi\hb
\eel{e:42}
where all ghost dependence has been eliminated. The quantization rule 1 allows
for two possibilities:

{\sl Case 1}:\hspace{1cm} $X^0, P^0$ has real spectra and $\pi, v$ imaginary.
This implies
\be
&&_{\pm}\vb ph|ph\hb_{\pm}\propto\int d^4p du e^{\pm
i(p^2+m^2)u}|\phi(p)|^2=2\pi\int
d^4\del(p^2+m^2)|\phi(p)|^2>0\nn\\
 \eel{e:43}

{\sl Case 2}:\hspace{1cm} $X^0, P^0$ has imaginary spectra and $\pi, v$ realy.
This implies
\be
&&_{\pm}\vb ph|ph\hb_{\pm}=\int d^4p dv e^{\pm
(p^2+m^2)v}\phi^*(-p^0,\bp)\phi(p^0, \bp)=\nn\\
&&=|\mbox{ finite only if } v\in(0, \infty) \mbox{ or
}(-\infty, 0)\;| =\nn\\
&&=\int d^4p\frac{\phi^*(-p^0,\bp)\phi(p^0,\bp)}{p^2+m^2}
 \eel{e:44}
The last norm is only positive if $\phi(p)$ has even parity. A Lorentz
covariant way to assure
positivity is to impose invariance under strong reflection $p^{\mu}\ra
-p^{\mu}$ on the original
state space $\Om$. Notice that the measure $d^4p$ is Euclidean in \r{44}. The
Euclidean propagator
is of the form $\vb ix|e^{[\rho, Q]}|ix'\hb$.

The BRST formulation of the worldline supersymmetric free massless spin-$\halv$
particle involves a
BRST charge of the form
 \be
 &&Q=P^2\eta+P\cdot\ga
c+\pet c^2+\pi\bapet+\kappa\bak \eel{e:45}
where the variables satisfy the following (anti-)commutation relations (the
nonzero part):
\be
&&[\ga^{\mu},\ga^{\nu}]_+=-2\eta^{\mu\nu}, \;\;\; [X^{\mu},
P^{\nu}]_-=i\eta^{\mu\nu},
\;\;\;[\pi, v]_-=-i,\;\;\;[\kappa, \la]_+=1,\nn\\
&&[\pet, \eta]_+=1,\;\;\;[\bapet, \bata]_+=1,\;\;\;[k, c]_-=-i,\;\;\;[\bak,
\bac]_-=-i
\eel{e:46}
where $k,c$ are bosonic ghosts and $\bak, \bac$ the corresponding antighosts,
$\la$ is a fermionic
Lagrange multiplier and $\kappa$ its conjugate momentum. $\eta^{\mu\nu}$ is a
space-like Minkowski
 metric. Notice that
\be
&&[P\cdot\ga, P\cdot\ga]_+=-2P^2
\eel{e:47}
is the algebra of the world-line supersymmetry. In the matrix representation
$\ga^{\mu}$ is turned
into the ordinary Dirac gamma matrices. The BRST charge \r{45} may also be
written as
$Q=\del+\del\dagg$ where \cite{Propa}
\be
&&\del=a\dagg D+\sigma\dagg\phi, \;\;\;[D, D]_+=-2\phi
\eel{e:48}
where in turn
\be
&&a\equiv\halv(c-i\bak),\;\;\;\sigma\equiv\halv(\eta-i\bapet-i\la c),\nn\\
%% FOLLOWING LINE CANNOT BE BROKEN BEFORE 80 CHAR
&&D=P\cdot\zeta+\la\pi-i\kappa-a^{\dag}\omega-\omega^{\dag}a,\;\;\;\phi=P^2-i\pi
\eel{e:481}
A formal algebraic  solution is \cite{Propa}
\be
&&|ph\hb_{\pm}=e^{\pm[\rho, Q]}|\Phi\hb,\;\;\;\rho=\pet v+k\la
\eel{e:49}
where
\be
&&c|\Phi\hb=\eta|\Phi\hb=\pi|\Phi\hb=\kappa|\Phi\hb=0
\eel{e:50}
The formal norm is
\be
&&_{\pm}\vb
ph|ph\hb_{\pm}=\vb\Phi|e^{\pm2(P^2v+iP\cdot\ga\la+k\bak+i\pet\bapet-2i\la\pet
c)}|\Phi\hb
 \eel{e:51}
The  quantization rule 2 reduces this expression to
\be
&&_{\pm}\vb
ph|ph\hb_{\pm}\propto\,'\vb\Phi|e^{\pm2P^2v}P\cdot\ga\la\pet\bapet|\Phi\hb'
\eel{e:52}
where $|\Phi\hb'$ is equal to $|\Phi\hb$ without the bosonic ghost part. At
this point there are
again two cases:

{\sl Case 1}:\hspace{1cm} $X^0, P^0$ has real spectra and $\pi, v$ imaginary.
This implies
\be
&&_{\pm}\vb ph|ph\hb_{\pm}\propto\int
d^4\del(p^2)\bar{\psi}(p){p}\!\!\!\slash\ga^5\psi(p)
 \eel{e:53}
which is not positive without further projections.

{\sl Case 2}:\hspace{1cm} $X^0, P^0$ has imaginary spectra and $\pi, v$ real.
This implies
\be
&&_{\pm}\vb ph|ph\hb_{\pm}\propto\int d^4p \int
d^4p\bar{\psi}(-p^0,\bp)\frac{{p}\!\!\!\slash\ga^5}{p^2+m^2}\psi(p^0, \bp)
 \eel{e:54}
which also is not positive without further projections.

There is an $O(2)$ extended worldline supersymmetric model for a massless
spin-1 particle
\cite{Otwo,MU}. Its gauge generators are
\be
&&P^2,\;\;\; P\cdot\ga_1,\;\;\;P\cdot\ga_2 ,\;\;\;\ga_1\cdot\ga_2
\eel{e:55}
where $\ga_i$ satisfies $[\ga_1, \ga_2]_+=0$ and $[\ga_i,
\ga_i]_+=-2\eta^{\mu\nu}$.
This is a noncanonical theory since
there exists no gauge fixing to $\ga_1\cdot\ga_2$. Within a BRST quantization
on inner product
spaces this implies that the physical state space will contain ghost
excitations and will not be
positive. A remedy of this defect is   to first restrict the original state
space by a condition of
the form  $\ga_1\cdot\ga_2\bar{\Om}=0$. A BRST treatment as above \cite{Propa}
yields then in case 1:
$\vb ph|ph\hb\propto\int d^4p\del(p^2)A^2_T(p)>0$. Case 2 yields on the other
hand Euclidean norms and
propagators in agreement with the path integral treatment in \cite{PRi}.

Above we have demonstrated that we are able to check unitarity of models for
relativistic particles
with our general solutions, \ie we can explicitly check whether or not  the
physical norms are
positive. Another issue raised in the introduction concerned the relation
between  path integrals
and the operator quantization. This relation turns out to be nontrivial
\cite{Path}. To see this I
consider the time evolution of the states \r{21}. Let $H_0$ be a BRST invariant
Hamiltonian. I
have then ($[\rho, H_0]$ is assumed to be BRST invariant)
\be
&&\vb ph',t'|ph,t\hb=\vb ph'|e^{-i(t'-t)H_0}|ph\hb=\nn\\
&&=\vb \Phi'|e^{\al[\rho,
Q]}e^{-i(t'-t)H_0}e^{\al[\rho, Q]}|\Phi\hb=\vb \Phi'|e^{-i(t'-t)H_0+2\al[\rho,
Q]}|\Phi\hb
\eel{e:56}
from \r{21}. $\al$ is a real constant different from zero. Since \r{56} is
independent of the value of
$\al$ I may set
\be
&&2\al=\pm(t'-t)
\eel{e:57}
for $t'\neq t$. Eqn \r{56} may then be interpreted as if the time evolution of
$|\Phi\hb$ is in terms
of a complex Hamiltonian. The conventional path integral expressions are still
obtained if $H_0$ is
represented by a real function and $[\rho, Q]$ by an imaginary one, a condition
which then
governs the possible choices of the quantization of the system. The latter
condition requires
the Lagrange multipliers to be quantized in consistency with an imaginary
spectral decomposition
which is in agreement with the quantization rule 1 above. Bosonic ghosts and
antighosts must be
quantized with opposite metric states in agreement with rule 2. For fermionic
ghosts and antighosts
one must choose real and imaginary odd Grassmann spectra respectively or vice
versa, a choice one
always may do  without affecting the metric of the states. Notice also that
fermionic
Lagrange multipliers must be quantized with imaginary odd spectra. If one uses
real spectra for
bosonic Lagrange multipliers, which was proposed as a possible choice in rule
1, then one finds a
complex Hamiltonian also in the path integrals except for the trivial case.
Thus, the Euclidean
treatment of the relativistic particles above corresponds to complex
Hamiltonians in the path
integrals. However, when they are analytically continued to the Minkowski space
they corresponds to
real Hamiltonians {\em apart from the $i\epsilon$-prescription in the
propagators}. It is this last
case that one usually makes use of in the path integral treatments
\cite{Teit,PRi,Gov} ($i\epsilon$ is then
introduced for convergence reason) although a treatment according to case 1
seems more natural
mathematically. Finally one may notice that the two cases of the general
solutions \r{21} yield
a precise connection between the choice of gauge fixing function $\rho$ and the
imposed boundary
conditions in the path integral a connection which is essentially in agreement
with those normally
made \cite{MH}.


\begin{thebibliography}{Fierz}

\bibitem{MU}R. Marnelius and U. M\aa rtensson, \  {\sl Nucl. Phys.}\
{\bf B321}, 185 (1989)


\bibitem{Teit}C. Teitelboim, \ {\sl Phys. Rev.}\ {\bf D25}, 3159 (1982); M.
Henneaux and C.
Teitelboim, \  {\sl Ann. Phys.}\ {\bf 143}, 127 (1982)

\bibitem{PRi}M. Pierri and V. O. Rivelles, \ {\sl  Phys. Lett.}\ {\bf 251B},
421 (1990)


\bibitem{Gov}J. Govaerts, \ {\sl Int. J. Mod. Phys.} \  {\bf 4}, 4487 (1989)

\bibitem{HM}S. Hwang
and R. Marnelius, \  {\sl Nucl. Phys.}\ {\bf B315}, 638
(1989);\ {\em
ibid.}\  {\bf B320}, 476 (1989)


\bibitem{MO}R. Marnelius and M. \"{O}gren, \
{\sl Nucl. Phys.}\ {\bf B351}, \ 474 \ (1991)

\bibitem{Bi}R. Marnelius, \
{\sl Nucl. Phys.}\ {\bf B370}, 165 (1992)



\bibitem{BV}I. A. Batalin and G. A. Vilkovisky, \
{\sl Phys. Lett.}
\ {\bf 69B},
309 (1977)


\bibitem{Simple}R. Marnelius, \ {\sl Nucl. Phys.}\ {\bf B395}, 647 (1993)

\bibitem{Gauge}R. Marnelius,
\  {\em Gauge fixing and abelianization in simple BRST quantization.},\
ITP-G\"{o}teborg report
93-17 (1993)


\bibitem{Aux}R. Marnelius, \ {\sl Nucl. Phys.}\ {\bf B372}, 218 (1992);\ {\em
ibid.}\  {\bf B384}, 318 (1992)

\bibitem{Propa}R. Marnelius,
\  {\em Proper BRST quantization of relativistic particles.}\ ITP-G\"{o}teborg
report 93-18 (1993)


\bibitem{Slav}S. A. Frolov and A. A. Slavnov, \ {\sl  Phys. Lett.}\ {\bf 218B},
461 (1989)

\bibitem{Pauli}W. Pauli, \  {\sl Rev. Mod. Phys.}\ {\bf 15}, 175
(1943)

\bibitem{Gen}R. Marnelius,
\  {\sl Nucl. Phys.}\ {\bf B391}, \ 621  \ (1993)



\bibitem{Otwo}V. D. Gershun and V. I. Tkach, \
{\sl JETP Lett.}\ {\bf 29},\ 288\ (1979);\\
R. Marnelius and B. Nilsson, \
{\sl ITP-G\"{o}teborg preprint}\ {79-52},\ (1979) (unpublished);\\
P. Howe, S. Penati, M. Pernici and P. Townsend\
{\sl Phys. Lett.}\ {\bf B215},\ 555\ (1988)


\bibitem{Path}R. Marnelius,
\  {\em A note on path integrals and time evolutions in BRST quantization.}\
ITP-G\"{o}teborg report 93-19 (1993)


\bibitem{MH}M. Henneaux, \  {\sl  Phys. Rep.}\ {\bf 126}, 1
(1985)



\end{thebibliography}
\end{document}